\documentclass[aps,pra,twocolumn,preprintnumbers,showpacs,superscriptaddress,amsmath,amssymb]{revtex4}
\usepackage{dcolumn}
\usepackage{bm}
\usepackage{amssymb}
\usepackage[german, english]{babel}
\usepackage{graphicx}
\usepackage{color}
\usepackage[letterpaper,total={7in,9.5in},top=0.75in,left=0.75in]{geometry}

\begin{document}
\title{Detection and manipulation of nuclear spin states in fermionic strontium}

\author{Simon Stellmer}
 \affiliation{Institut f\"ur Quantenoptik und Quanteninformation (IQOQI),
\"Osterreichische Akademie der Wissenschaften, 6020 Innsbruck,
Austria}
\affiliation{Institut f\"ur Experimentalphysik und
Zentrum f\"ur Quantenphysik, Universit\"at Innsbruck,
6020 Innsbruck, Austria}
\author{Rudolf Grimm}
 \affiliation{Institut f\"ur Quantenoptik und Quanteninformation (IQOQI),
\"Osterreichische Akademie der Wissenschaften, 6020 Innsbruck,
Austria}
\affiliation{Institut f\"ur Experimentalphysik und
Zentrum f\"ur Quantenphysik, Universit\"at Innsbruck,
6020 Innsbruck, Austria}
\author{Florian Schreck}
\affiliation{Institut f\"ur Quantenoptik und Quanteninformation (IQOQI),
\"Osterreichische Akademie der Wissenschaften, 6020 Innsbruck,
Austria}

\date{\today}


\begin{abstract}
Fermionic $^{87}$Sr has a nuclear spin of $I=9/2$, higher than any other element with similar electronic structure. This large nuclear spin has many applications in quantum simulation and computation, for which preparation and detection of the spin state are requirements. For an ultracold $^{87}$Sr cloud, we show two complementary methods to characterize the spin-state mixture: optical Stern-Gerlach state separation and state-selective absorption imaging. We use these methods to optimize the preparation of a variety of spin-state mixtures by optical pumping and to measure an upper bound of the $^{87}$Sr spin relaxation rate.
\end{abstract}

\maketitle

\section{Introduction}

Fermions with two valence electrons, like $^{43}$Ca, $^{87}$Sr, $^{171}$Yb, and $^{173}$Yb, have a rich internal state structure, which is at the heart of recent proposals for quantum simulation and computation \cite{Cazalilla2009ugo,Gorshkov2010tos,Wu2003ess,Wu2006hsa,FossFeig2010ptk,Hermele2009mio,Xu2010lim,Hung2011qmo,Gerbier2009gff,Cooper2011ofl,Beri2011zti,Gorecka2011smf,Daley2008qcw,Gorshkov2009aem}. Unlike bosonic isotopes of these elements, the fermions have a nuclear spin, which decouples from the electronic degrees of freedom in the $^1S_0$ ground state and the $^3P_0$ metastable state. This gives rise to a $SU(N)$ spin symmetry, where $N$ is the number of nuclear spin states, which can be as high as 10 for $^{87}$Sr \cite{Cazalilla2009ugo,Gorshkov2010tos}. Rich quantum phases have been predicted to exist in such Fermi systems
\cite{Wu2003ess,Wu2006hsa,Cazalilla2009ugo,FossFeig2010ptk,Hermele2009mio,Gorshkov2010tos,Xu2010lim,Hung2011qmo}. The nuclear spin is also essential for the implementation of artificial non-abelian gauge fields \cite{Gerbier2009gff,Beri2011zti,Gorecka2011smf}. Furthermore it can be used to robustly store quantum information, which can be manipulated using the electronic structure \cite{Daley2008qcw,Gorshkov2009aem}. After the recent attainment of quantum degeneracy in $^{171,173}$Yb \cite{Fukuhara2007dfg,Taie2010roa} and $^{87}$Sr \cite{DeSalvo2010dfg,Tey2010ddb}, these ideas are coming closer to realization.

Essential tools for quantum simulation and computation with these degenerate gases are the detection and manipulation of the spin-state mixture. Several alkaline-earth spin-state detection schemes were already demonstrated. The number of atoms in the highest $m_F$ state was determined by selectively cooling \cite{Mukaiyama2003rll} or levitating \cite{Tey2010ddb} atoms in this state. The number of atoms in an arbitrary $m_F$ state was determined using state-selective shelving of atoms in a metastable state, which requires a ``clock''-transition laser \cite{Boyd2007nse}. Recording the full $m_F$-state distribution with this method is possible, but needs one experimental run per state. Determination of the $m_F$-state distribution in only two experimental runs was recently shown for quantum-degenerate Yb gases, using optical Stern-Gerlach (OSG) separation \cite{Taie2010roa}.

In this Article, we first show two nuclear spin-state detection schemes for $^{87}$Sr that have advantages over the previously used schemes. In Sec.~\ref{sec:detection}, we present an adaptation and extension of the OSG separation scheme of \cite{Taie2010roa}, which in our implementation is able to resolve all ten nuclear spin-states of Sr in a single experimental run. In Sec.~\ref{sec:AbsorptionImaging}, we describe state-selective absorption imaging using the intercombination line of Sr. Unlike OSG separation, this method gives spatial information about the spin-state distribution and is also applicable to samples before evaporative cooling. In Sec.~\ref{sec:OpticalPumping}, we demonstrate the preparation of a desired spin-state mixture by optical pumping, using spin-state detection to optimize the optical pumping procedure. In Sec.~\ref{sec:SpinRelaxation}, we determine an upper limit of the $^{87}$Sr spin-relaxation rate, using our spin-state preparation and detection methods.

\section{Optical Stern-Gerlach separation}
\label{sec:detection}

Here, we describe the operation principle of optical Stern-Gerlach (OSG) separation (Sec.~\ref{sec:OSGOperationPrinciple}) and our experimental implementation (Sec.~\ref{sec:OSGExperimentalDemonstration}). In Sec.~\ref{sec:OSGSimulation}, we describe a simulation of the OSG process. In Sec.~\ref{sec:OSGAtomnumberDistribution} we determine the number of atoms in each spin state.

\subsection{Operation principle}
\label{sec:OSGOperationPrinciple}

The Stern-Gerlach technique separates atoms in different internal states by applying a state-dependent force and letting the atomic density distribution evolve under this force \cite{Stern1922den}. The implementation of this technique for alkali atoms is simple. Their single valence electron provides them with a $m_F$-state dependent magnetic moment that, for easily achievable magnetic field gradients, results in $m_F$-state dependent forces sufficient for state separation \cite{StamperKurn1998oco}. By contrast, atoms with two valence electrons possess only a weak, nuclear magnetic moment in the electronic ground state, which would require the application of impractically steep magnetic field gradients. An alternative is OSG separation, where a state dependent dipole force is used. OSG separation was first shown for a beam of metastable helium \cite{Sleator1992edo}, where orthogonal dressed states of the atoms were separated by a resonant laser field gradient. The case of interest here, OSG $m_F$-state separation, was recently realized for a quantum degenerate gas of Yb, by using $m_F$-state dependent dipole forces \cite{Taie2010roa}.

\begin{figure}
\includegraphics[width=\columnwidth]{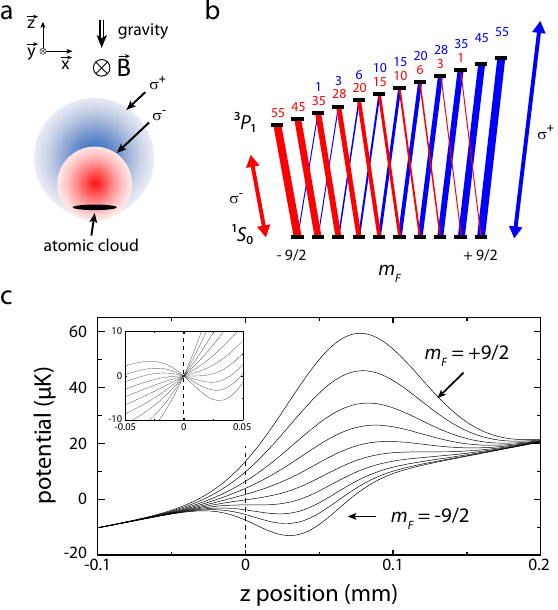}
\caption{\label{fig:Fig1} (Color online) Principle of OSG separation. a) $\sigma^{+}$- and $\sigma^{-}$-polarized laser beams propagating in the $y$-direction create dipole forces on an atomic cloud that is located on the slopes of the Gaussian beams. b) The laser beams are tuned close to the $^1S_0$($F=9/2$) - $^3P_1$($F'=11/2$) intercombination line, creating attractive ($\sigma^{-}$ beam) or repulsive ($\sigma^{+}$ beam) dipole potentials. Each $m_F$ state experiences a different potential because of the varying line strength of the respective transition. c) The potentials resulting from dipole potentials and the gravitational potential. The dashed line marks the initial position of the atoms. The inset shows the relevant region of the potentials, offset shifted to coincide at the position of the atoms, which clearly shows the different gradient on each $m_F$ state.}
\end{figure}

We first explain the basic operation principle of Sr OSG separation before discussing our experimental implementation. The experimental situation is shown in Fig.~\ref{fig:Fig1}(a). An ultracold cloud of $^{87}$Sr atoms in a mixture of $m_F$ states is released from an optical dipole trap. The $m_F$-state dependent force is the dipole force of two laser beams propagating in the plane of the pancake-shaped cloud, one polarized $\sigma^{+}$, the other $\sigma^{-}$. The diameter of these OSG laser beams is on the order of the diameter of the cloud in the $x$-direction. The beams are displaced vertically by about half a beam radius to produce a force in the $z$-direction on the atoms. To create a $m_F$-state dependent force, the OSG beams are tuned close to the $^1S_0$($F=9/2$) - $^3P_1$($F'=11/2$) intercombination line (wavelength 689\,nm, linewidth 7.6\,kHz), so that this line gives the dominant contribution to the dipole force. A guiding magnetic field is applied in the direction of the laser beams such that the beams couple only to $\sigma^{+}$ or $\sigma^{-}$ transitions, respectively. The line strength of these transitions varies greatly with the $m_F$ state \cite{Metcalf1999book}, see Fig.~\ref{fig:Fig1}(b), resulting in different forces on the states. For $^{173}$Yb, this variation, together with a beneficial summation of dipole forces from transitions to different $^3P_1$ hyperfine states, was sufficient to separate four of the six $m_F$ states using just one OSG beam \cite{Taie2010roa}. The remaining two $m_F$ states could be analyzed by repeating the experiment with opposite circular polarization of the OSG beam.

Strontium, which has nearly twice as many nuclear spin states, requires an improved OSG technique to separate the states. The improvement consists of applying two OSG beams with opposite circular polarization at the same time. The $\sigma^{+}$-polarized beam produces dipole forces mainly on the positive $m_F$ states, the $\sigma^{-}$ beam mainly on the negative $m_F$ states. By positioning the beams in the appropriate way (see below), the forces point in opposite directions and all $m_F$ states can be separated in a single experimental run. A second improvement is to enhance the difference in the dipole forces on neighboring $m_F$ states by tuning already strong transitions closer to the OSG beam frequency using a magnetic field, which splits the excited state $m_{F'}$ states in energy. For our settings, the difference in forces on neighboring high $|m_F|$ states is enhanced by up to 25\%, which helps to separate those states. This enhancement scheme requires the $\sigma^{+}$-polarized OSG beam to be tuned to the blue of the resonance, whereas the $\sigma^{-}$ beam has to be tuned to the red of the resonance, see Fig.~\ref{fig:Fig1}(b). Both beams are centered above the atomic cloud so that the repulsive blue detuned beam produces a force pointing downwards, whereas the attractive red detuned beam produces a force pointing upwards.

\begin{figure}
\includegraphics[width=\columnwidth]{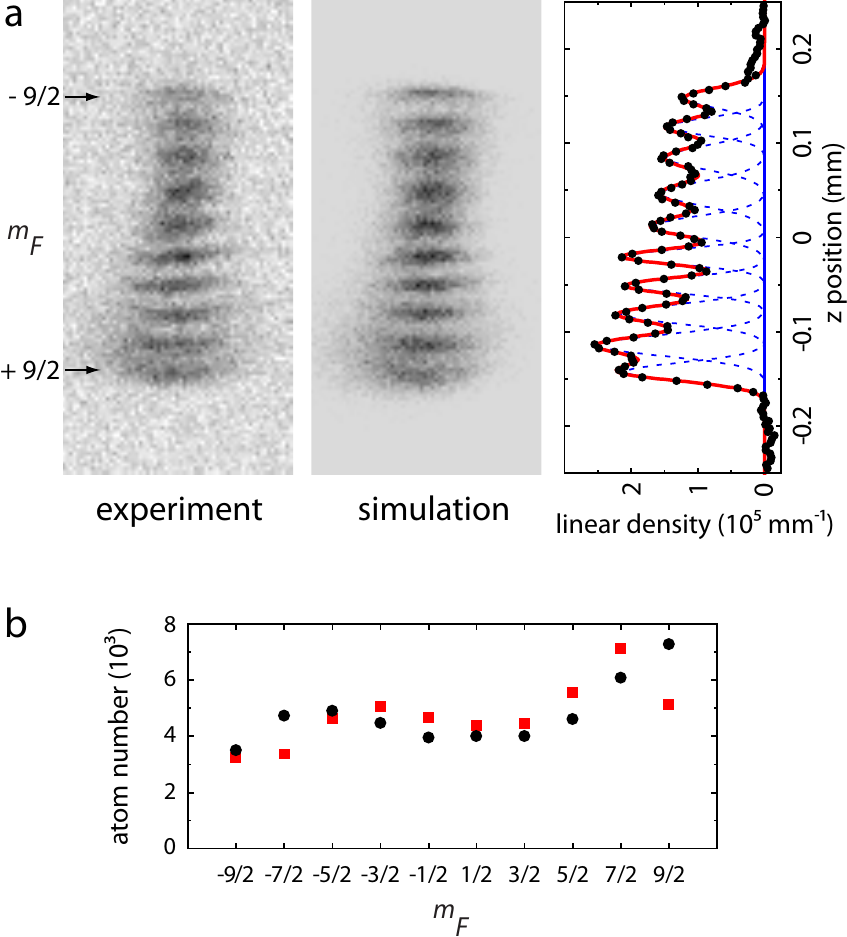}
\caption{\label{fig:Fig2} (Color online) OSG separation of the ten $^{87}$Sr nuclear spin states. a) Atomic density distribution after OSG separation integrated over the $(\mathbf{\hat{x}}+\mathbf{\hat{y}})$-direction obtained in experiment and simulation. On the right, the density distribution of the experiment integrated along the $x$- and $y$-directions is shown together with a fit consisting of ten Gaussian distributions. b) Atom number distribution determined from the area of the Gaussian distributions (squares) and the fit of the simulation to the experiment (circles).}
\end{figure}

\subsection{Experimental demonstration}
\label{sec:OSGExperimentalDemonstration}

We demonstrate OSG separation of a cloud of $4.5 \times 10^4$ $^{87}$Sr atoms in a mixture of $m_F$ states. To prepare the cloud, Zeeman slowed $^{87}$Sr atoms are laser cooled in two stages, first in a ``blue'' magneto-optical trap (MOT) on the broad-linewidth $^1S_0$-$^1P_1$ transition, then in a ``red'' MOT on the narrow-linewidth $^1S_0$-$^3P_1$ transition \cite{DeSalvo2010dfg,Tey2010ddb}. Next, the atoms are transferred to a pancake-shaped optical dipole trap with strong confinement in the vertical direction. The sample is evaporatively cooled over seven seconds. At the end of evaporation the trap oscillation frequencies are $f_x=19\,$Hz, $f_y=11$\,Hz, and $f_z=85$\,Hz, where the coordinate system is defined in Fig.~\ref{fig:Fig1}(a). The collision rate at this stage is only $1\,$s$^{-1}$, which is insufficient for complete thermalization. Since atoms are evaporated mainly downwards, along the $z$-direction, the sample is not in cross-dimensional thermal equilibrium, having a temperature of 25\,nK in the $z$-direction and twice that value in the $xy$-plane. The sample is non-degenerate and the $1/e$-widths of the Gaussian density distribution are $w_x=55\,\mu$m, $w_y=85\,\mu$m, and $w_z=7\,\mu$m.

The OSG beams propagate along the $y$-direction. The power of the $\sigma^{+}$ ($\sigma^{-}$) beam is 4\,mW (0.5\,mW), the waist is $\sim$80\,$\mu$m ($\sim$60\,$\mu$m), and the beam center is displaced $\sim$70\,$\mu$m ($\sim$40\,$\mu$m) above the cloud. Both beams create dipole forces of similar magnitude since the reduced power of the $\sigma^{-}$ beam compared to the $\sigma^+$ beam is partially compensated by its decreased waist. At zero magnetic field, the $\sigma^{\pm}$ beam is detuned $\pm100\,$MHz from resonance. To increase the difference in dipole potential on neighboring $m_F$ states, a magnetic field of 16\,G is applied parallel to the OSG beams, which splits neighboring $^3P_1$($F'=11/2$) $m_{F'}$ states by 6.1\,MHz. With this field applied, the $\sigma^{\pm}$ beam has a detuning of $\pm 66.4$\,MHz to the $^1S_0$($F=9/2$, $m_F=\pm9/2$) - $^3P_1$($F'=11/2$, $m_F=\pm11/2$) transition and a detuning of $\pm133.6$\,MHz to the $^1S_0$($F=9/2$, $m_F=\mp9/2$) - $^3P_1$($F'=11/2$, $m_F=\mp7/2$) transition, see Fig.~\ref{fig:Fig1}(b).

OSG separation is started by simultaneously releasing the atoms from the dipole trap and switching on the OSG beams. The atoms are accelerated for 1.6\,ms by the OSG beams. Then the beams are switched off to avoid oscillations of atoms in the dipole trap formed by the red detuned OSG beam. The atoms freely expand for another 2.3\,ms before an absorption image on the $^1S_0$-$^1P_1$ transition is taken. The result is shown in Fig.~\ref{fig:Fig2}(a). All ten $m_F$ states are clearly distinguishable from each other.

To obtain a good separation of the $m_F$ states and an even spacing between them, OSG beam waists, the timing of the OSG separation sequence, the applied magnetic field, and the beam positions were optimized. We found that for all other parameters fixed, the position of the OSG beams is critical and has to be aligned to better than 10\,$\mu$m.

To quantify the separation of the states, we fit ten Gaussian distributions to the density distribution integrated along the $x$- and $y$-directions, see Fig.~\ref{fig:Fig2}(a). We obtain a separation of adjacent states between 28 and 38\,$\mu$m, similar to the $1/e$-widths of the distributions, which are between 24 and 36\,$\mu$m. The $1/e$-width expected from initial size and temperature after 3.9\,ms total expansion time is 19\,$\mu$m in the $z$-direction, slightly narrower than the width of the narrowest distributions observed. From the Gaussian fits we also obtain an estimation of the atom number in each state, see Sec.~\ref{sec:OSGAtomnumberDistribution}.

OSG separation works only well for very cold samples. If the temperature is too high, the sample expands too fast and the individual $m_F$-state distributions cannot be distinguished. For a density minimum to exist between two neighboring $m_F$-state distributions of Gaussian shape, the $1/e$-widths have to be smaller than $\sqrt{2}$ times the distance between the maxima of the distributions. For our smallest separation of 24\,$\mu$m, this condition corresponds to samples with a temperature below 100\,nK, which can only be obtained by evaporative cooling.

\subsection{Simulation}
\label{sec:OSGSimulation}

\begin{figure}
\includegraphics[width=\columnwidth]{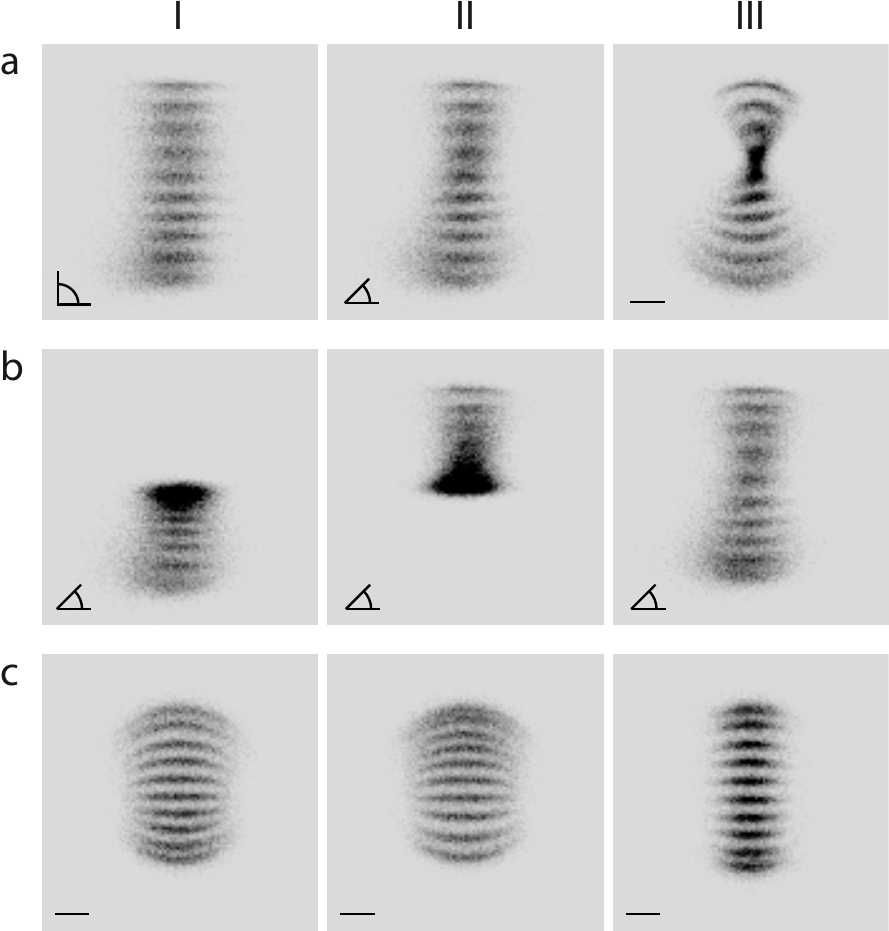}
\caption{\label{fig:Fig3} Simulated density distributions of atoms after OSG separation. Shown are distributions integrated along $x$ (aI), along the $(\mathbf{\hat{x}}+\mathbf{\hat{y}})$-direction (aII,b), and along $y$ (aIII,c). The angle symbols indicate the angle between the direction of integration and the direction of the OSG beams, the $y$-direction. a) The density distribution obtained from a simulation with parameters fitted to the outcome of the experiment, integrated along different directions. The experimental situation corresponds to (aII). b) Influence of the two OSG beams and the magnetic field on OSG separation. bI) As (aII), but no $\sigma^{-}$ beam. bII) As (aII), but no $\sigma^{+}$ beam. bIII) As (aII), but no magnetic field and reduced OSG beam detuning. c) Alternative scheme of OSG separation using two blue detuned OSG beams and variations of that scheme, see main text.}
\end{figure}

We perform a simulation of classical trajectories of atoms to better understand the OSG separation process. The simulation takes into account the dipole potentials of the OSG beams, discrete spontaneous scattering events of OSG beam photons, and gravity. The phase-space distribution of simulated atoms is initialized using the experimentally measured trap frequencies and temperatures. The calculated density distributions are fit to the experimental result using the OSG beam waists and positions and the atom number of each state as fit parameters. The detunings and intensities of the OSG beams and the value of the magnetic field are fixed to the values used in the experiment \cite{EndnoteOSGBeamIntensityVariation}. The parameters of the $\sigma^{+}$ ($\sigma^{-}$) beam resulting in the best fit are a waist of 90\,$\mu$m (56\,$\mu$m) and a displacement of the beam center relative to the atomic cloud's center by 74\,$\mu$m (35\,$\mu$m) in the $z$-direction and by 0\,$\mu$m (7\,$\mu$m) in the $x$-direction. With these parameters, the simulation matches the experimental result well, see Fig.~\ref{fig:Fig2}(a).

We now analyze the OSG separation process in detail using the simulation. The OSG beam potential gradients on atoms in different $m_F$ states are nearly evenly distributed between 47\,m/s$^2$ downwards for the $+9/2$ state and 32\,m/s$^2$ upwards for the $-9/2$ state (see inset of Fig.~\ref{fig:Fig1}(c)), resulting in the nearly evenly spaced $m_F$-state distribution after OSG separation. About 25\% of the atoms spontaneously scatter photons from the OSG beams, which leads to a widening of the individual $m_F$-state distributions in the $z$-direction by at most 20\%, which only insignificantly reduces our ability to distinguish the $m_F$ states.

Also the $^3P_1$ $F'=7/2$ and $F'=9/2$ hyperfine levels contribute noticeably to the dipole forces created by the OSG beams on the low $|m_F|$ states, which at first glance might be surprising since the detuning of the OSG beams from those states is more than an order of magnitude larger than the detuning from the $F'=11/2$ state. The reason is that the dipole forces from the $\sigma^{+}$ and $\sigma^{-}$ OSG beams considering only the $F'=11/2$ hyperfine level nearly compensate each other for low $|m_F|$ states, whereas the dipole forces from both beams considering the other hyperfine levels point into the same direction and in addition are strongest for low $|m_F|$ states. The positions of the low $|m_F|$ states after a simulated OSG separation with and without considering the influence of the $F'=7/2$ and $F'=9/2$ levels changes by up to 13\,$\mu$m. By contrast, the positions of the $m_F=\pm9/2$ states are changed by less than 2\,$\mu$m. It is not possible to obtain a good fit of the simulation to the experiment if the influence of the $^3P_1$ $F'=7/2$ and $F'=9/2$ hyperfine levels is neglected.

The shape of the density distribution is analyzed in row (a) of Fig.~\ref{fig:Fig3}. The density distribution is shown as it would appear using absorption imaging along the $x$- (aI), the $(\mathbf{\hat{x}}+\mathbf{\hat{y}})$- (aII), or the $y$-direction (aIII). Case (aII) is the one realized in the experiment. Strong distortions of the spatial distribution of each $m_F$ state compared to a free expansion are visible. They are induced by the finite size of the OSG beams. The blue detuned beam expels high $m_F$-state atoms onto cylindrical surfaces, whereas the red detuned beam attracts low $m_F$-state atoms and acts like a focussing lens.

The role of each OSG beam and the magnetic field are demonstrated in row (b) of Fig.~\ref{fig:Fig3}. Shown are the results of simulations with only the blue (bI) or the red (bII) detuned OSG beam present. With only one OSG beam, at best four $m_F$ states can be well separated, illustrating the need of two beams for Sr OSG separation. In simulation (bIII) the magnetic field was set to zero, which removes the energy splitting of the $^3P_1$($F'=11/2$) manifold. To achieve the same acceleration on the $m_F=\pm9/2$ states as with magnetic field, the detuning of the $\sigma^{\pm}$ OSG beam was set to $\pm66.4\,$MHz from the center of the $F'=11/2$ manifold. Without magnetic field, the spatial splitting between neighboring high $|m_F|$ states is slightly reduced. To separate these states further than done in the experiment, a larger magnetic field could be used.

The simulation also suggests an alternative OSG separation scheme, which is demonstrated in row (c) of Fig.~\ref{fig:Fig3}, but which we did not check experimentally. Instead of one blue and one red detuned OSG beam, the scheme uses two blue detuned OSG beams of opposite circular polarization. As before, the $\sigma^{+}$ beam is placed above the cloud, acts mainly on the high $m_F$-state atoms and pushes them downwards. The $\sigma^-$ beam uses now the same detuning, power, and waist as the $\sigma^+$ beam, but is placed below the initial center of the atomic cloud. The location of the two beams in the $z$-direction is symmetric with respect to the initial position of the atomic cloud. The $\sigma^-$ beam acts mainly on the negative $m_F$-state atoms and pushes them upwards. Since this beam is now blue detuned, it does not act similar to a lens as the red detuned $\sigma^-$ beam used in the experiment and leads to less distortion of the cloud. Three cases of this alternative scheme are shown in row (c), always assuming an even atom number distribution over the $m_F$ states. Case (cI) uses $\sigma^{+}$-beam parameters equivalent to the ones used for the simulations of row (a) and $\sigma^-$-beam parameters deduced from those as described above. In addition the magnetic field is set to zero. This situation leads to a nearly symmetric separation of positive and negative $m_F$ states, where the symmetry is only slightly broken by gravity. Compared to the situation realized in the experiment, atoms in low $m_F$ states are better separated.

One slight drawback of this scheme is that the application of a magnetic field as used in the experiment will only increase separation of positive $m_F$ states. The separation of negative $m_F$ states will even be decreased. This effect is demonstrated in case (cII), where a magnetic field similar to the one used in the experiment is assumed and the detunings of the OSG beams changed such that the initial accelerations of the $m_F=\pm9/2$ states are the same as in case (cI). The reason for the decreased separation of negative $m_F$ states is a reduction of the difference in the dipole forces of the $\sigma^-$ beam on neighboring $m_F$ states. This reduction comes from the Zeeman splitting of the $^3P_1$ $F'=11/2$ level, which will tune transitions with strong line strength farther away from the $\sigma^-$-beam wavelength than transitions with weak linestrength.

The distortions of the density distribution after OSG separation can be reduced if more power is available for the OSG beams. Then the waist of the OSG beams can be made wider in the $x$-direction keeping the same potential in the $z$-direction. This leads to a reduction of unwanted potential gradients along $x$, which are the source of the distortions well visible e.g. in (aIII) or (cI). The reduction is demonstrated in case (cIII), where the power and waist in the $x$-direction of the OSG beams has been doubled compared to case (cI).

\subsection{Determination of the atom number distribution}
\label{sec:OSGAtomnumberDistribution}

To obtain the number of atoms in each $m_F$ state, we use two approaches. The first one determines the atom numbers from ten Gaussian fits to the density distribution integrated along the $x$- and $y$-directions, see Fig.~\ref{fig:Fig2}(a). The second approach uses the fit of the simulation to the data described above, which takes the distortions of the distribution better into account, but relies on our ability to accurately model the OSG separation process. A systematic effect should be considered in both approaches: the $m_F$-state dependence of the average photon number absorbed by an atom during absorption imaging. This dependence has its origin in the $m_F$-state dependent line strength of the absorption imaging transition. Under our imaging conditions (probe beam of circularly polarized light on the $^1S_0$-$^1P_1$ transition with an intensity of 0.5\,mW/cm$^2$, an angle of 45$^\circ$ to the quantization axis given by the magnetic field, and 40\,$\mu$s exposure time) in average about 40 photons are scattered per atom, making optical pumping during absorption imaging important. We simulate this optical pumping process to obtain an estimate of the number of photons scattered by an atom in dependence of its initial $m_F$ state. We find that the optical pumping process depends strongly on the detuning of the absorption imaging beam to the closely spaced hyperfine states of the $^1P_1$ excited state, which are mixed at the 16\,G magnetic field applied. Because of this dependence, not only the absolute number of photons scattered per atom depends on the detuning, but also the ratio of the number of photons scattered for different initial states. For atoms starting in the $m_F=+9/2$ or the $m_F=-9/2$ state the ratio is highest, about 1.2(2), where the error comes from the uncertainty of the laser detuning used in the experiment. Further experimental study of this effect is needed to determine the relative atom numbers better. Simply assuming equal and maximal absorption by atoms in each $m_F$ state, the atom number distributions resulting from the two approaches are shown in Fig.~\ref{fig:Fig2}(b). The atom numbers of both approaches agree to better than 20\% for all but two $m_F$ states. The agreement is less good for the $m_F=+9/2$ state, which has the most distorted distribution, and the $m_F=-7/2$ state, for which the Gaussian fit underestimates the width.

\section{Spin-state dependent absorption imaging}
\label{sec:AbsorptionImaging}

\begin{figure}
\includegraphics[width=\columnwidth]{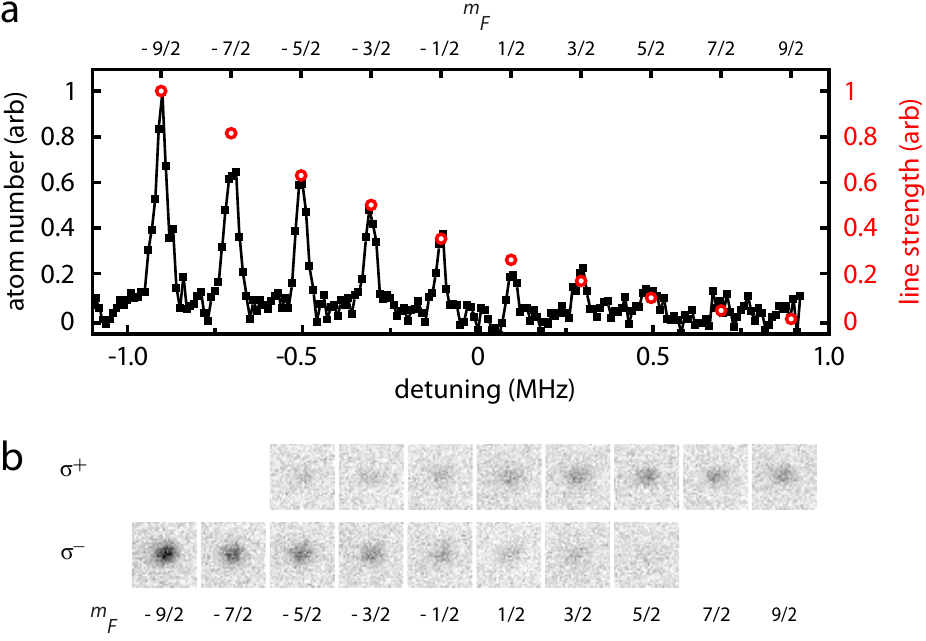}
\caption{\label{fig:Fig4} (Color online) $m_F$-state resolved absorption imaging on the $^1S_0$($F=9/2$) - $^3P_1$($F'=11/2$) intercombination line. a) Spectrum of a $^{87}$Sr sample with nearly homogeneous $m_F$-state distribution. The spectrum was obtained using $\sigma^{-}$-polarized light and shifting transitions corresponding to different $m_F$ states in frequency by applying a magnetic field of 0.5\,G. The circles give the line strengths of the transitions. b) Absorption images taken on the maxima of absorption of each $m_F$ state using $\sigma^{+}$ or $\sigma^{-}$ polarized light.}
\end{figure}

We also demonstrate a complementary method of $m_F$-state detection: $m_F$-state dependent absorption imaging. This method is often used for alkali atoms employing a broad linewidth transition \cite{Matthews1998dro}. For Sr, $m_F$-state resolved imaging on the broad $^1S_0$-$^1P_1$ transition is not possible since the magnetic field splitting of the exited state $m_{F'}$ states is smaller than the linewidth of the transition \cite{Boyd2007nse}. But $m_F$-state dependent imaging can be realized using the narrow $^1S_0$($F=9/2$) - $^3P_1$($F'=11/2$) intercombination line. To achieve state selectivity, we apply a magnetic field of 0.5\,G, which splits neighboring $m_{F'}$ states by 200\,kHz, which is 27 times more than the linewidth of the imaging transition. The advantages of this method compared to OSG separation is its applicability to samples that have not been evaporatively cooled, spatially resolved imaging, and a near perfect suppression of signal from undesired $m_F$ states. A disadvantage of this method is that it delivers a reduced signal compared to imaging on the $^1S_0$-$^1P_1$ transition, as done after OSG separation. The reduction comes from the narrower linewidth, optical pumping to dark states during imaging, and weak line strengths for some $m_F$ states. Figure~\ref{fig:Fig4} shows a spectroscopy scan and absorption images taken on the maxima of the absorption signal for a sample with nearly homogeneous $m_F$-state distribution. The absorption is strongly $m_F$-state dependent and to obtain the best signal, the polarization of the absorption imaging light has to be adapted to the $m_F$ state of interest: $\sigma^{+}$($\sigma^{-}$) for high (low) $m_F$ states and $\pi$ for low $|m_F|$ states. For our absorption imaging conditions (an intensity of 15\,$\mu$W/cm$^2$, which is 5 times the saturation intensity, and an exposure time of 40\,$\mu$s), even atoms in $m_F$ states corresponding to the strongest transition will on average scatter less than one photon. Therefore the absorption is expected to be nearly proportional to the $m_F$-state dependent line strength of the transition, which we confirm using a simulation of the absorption imaging process \cite{EndnoteAbsImagingSimulationDetails}.

\section{Preparation of spin-state mixtures}
\label{sec:OpticalPumping}

\begin{figure}
\includegraphics[width=\columnwidth]{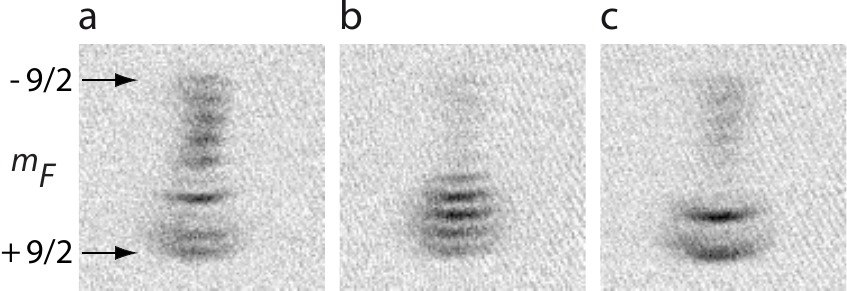}
\caption{\label{fig:Fig5} Examples of state mixtures prepared by optical pumping and analyzed using OSG separation. (a) The $m_F=1/2$ and $5/2$ states were pumped to the $3/2$ state using $\sigma^{+}$ and $\sigma^{-}$ light, respectively. (b) The negative $m_F$ states were pumped to the positive $m_F$ states. (c) A two-state mixture obtained by pumping the lower $m_F$ states to the $m_F=5/2$ and $7/2$ states and subsequently pumping the $m_F=7/2$ state to the $m_F=9/2$ state.}
\end{figure}

For applications of $^{87}$Sr to quantum simulation and computation, the $m_F$-state mixture needs to be controlled. We produce a variety of different mixtures by optical pumping, making use of OSG separation to quickly optimize the optical pumping scheme and quantify the result. Optical pumping is performed on the $^1S_0$($F=9/2$) -$^3P_1$($F'=9/2$) intercombination line, before evaporative cooling. A field of 3\,G splits neighboring excited state $m_{F'}$ states by 255\,kHz. This splitting is well beyond the linewidth of the transition of 7.4\,kHz, allowing transfer of atoms from specific $m_F$ states to neighboring states using $\sigma^{\pm}$- or $\pi$-polarized light, the choice depending on the desired state mixture. Sequences of pulses on different $m_F$ states can create a wide variety of state mixtures, of which three examples are shown in Fig.~\ref{fig:Fig5}. Optical pumping on the $^1S_0$($F=9/2$) - $^3P_1$($F'=7/2$) and $^1S_0$($F=9/2$) - $^3P_1$($F'=11/2$) transitions yields similar results.

\section{Determination of an upper bound of the spin-relaxation rate}
\label{sec:SpinRelaxation}

\begin{figure}
\includegraphics[width=\columnwidth]{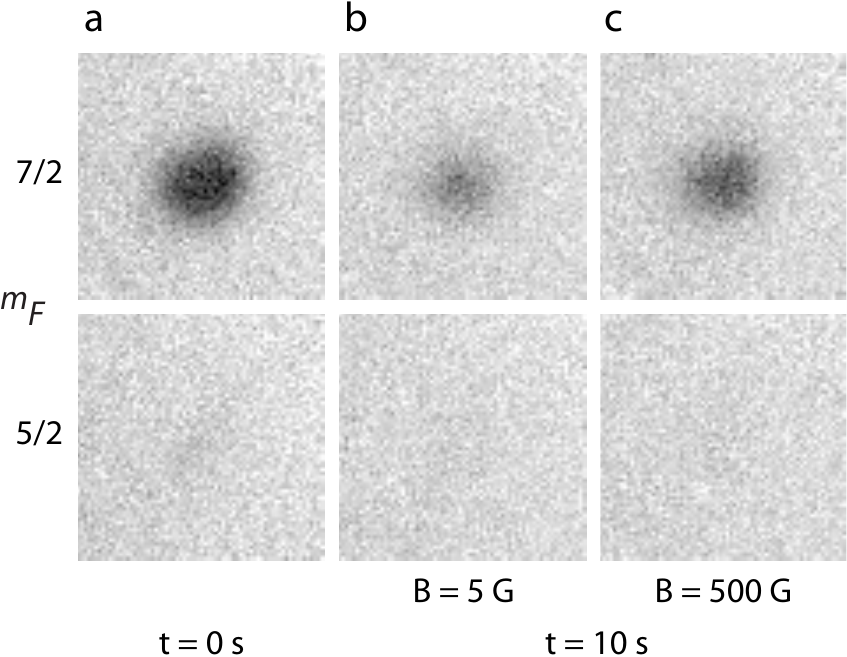}
\caption{\label{fig:Fig6} Absence of spin relaxation in $^{87}$Sr. Shown are absorption images of the $m_F=5/2$ and $7/2$ state averaged over 25 runs of the experiment. Atoms were initially removed from the $m_F=5/2$ state by optical pumping, whereas all other $m_F$ states remained populated (a). After 10\,s hold at a magnetic field of 5\,G (b) or 500\,G (c) no $m_F=5/2$ atoms are detectable, showing the low rate of spin relaxation.}
\end{figure}

A low nuclear spin-relaxation rate is an essential requirement to use $^{87}$Sr for quantum simulation and computation \cite{Cazalilla2009ugo,Gorshkov2010tos}. The rate is expected to be small since the nuclear spin does not couple to the electronic degrees of freedom in the ground state. Here, we use our nuclear spin state preparation and detection techniques to determine an upper bound for this spin relaxation rate. We start with a sample of $1.5\times 10^6$ atoms with near uniform $m_F$-state distribution and a temperature of $T=1.5\,\mu$K, confined in a trap with oscillation frequencies $f_x=67\,$Hz, $f_y=68\,$Hz, and $f_z=360\,$Hz, obtained directly after transferring the atoms from the magneto-optical trap to the dipole trap without any further evaporation. We optically pump all atoms from the $m_F=5/2$ state to neighboring states and look out for the reappearance of atoms in this state by spin relaxation during 10\,s of hold. The atom number in the $m_F=5/2$ state and, as a reference, the $m_F=7/2$ state are determined from absorption images. During 10\,s of hold at a magnetic field of either 5\,G or 500\,G the number of $m_F=5/2$ atoms remains below our detection threshold of about $10^4$ atoms, indicating a low spin-relaxation rate, see Fig.~\ref{fig:Fig6}. From this observation, we can obtain an upper bound for the spin-relaxation rate. To obtain a conservative bound, we assume that the dominant process leading to the creation of $m_F=5/2$-state atoms are collisions of $m_F=7/2$- with $m_F=3/2$-state atoms, forming two $m_F=5/2$-state atoms. Since the second order Zeeman effect is negligible no energy is released in such a collision and the resulting $m_F=5/2$-state atoms will remain trapped. The number of atoms created in the $m_F=5/2$ state by spin relaxation after a hold time $t$ is $N_{5/2}=2 N_{\rm state} g_{\rm sr} \overline{n} t$, where $N_{\rm state}=1.5 \times 10^5$ is the atom number in each populated state, $g_{\rm sr}$ the spin-relaxation rate constant, $\overline{n}=7.5 \times 10^{11}$\,cm$^{-3}$ the mean density and the factor 2 takes into account that two atoms are produced in the $m_F=5/2$ state per collision. From our measurement we know that $N_{5/2}<10^4$, from which we obtain an upper bound of $5 \times 10^{-15}\,$cm$^3$s$^{-1}$ for the spin-relaxation rate constant. This bound for the rate constant corresponds for our sample to a spin relaxation rate which is 2000 times smaller than the elastic scattering rate. The rate constant could be even orders of magnitude smaller than the already low upper bound we obtained \cite{JuliennePrivateComm}.

\section{Conclusion}
\label{sec:Conclusion}

In conclusion, we have demonstrated two complementary methods to characterize the spin-state mixture of an ultracold cloud of $^{87}$Sr. Optical Stern-Gerlach separation can characterize the spin-state mixture of an evaporatively cooled sample in a single experimental run and is very useful for fast optimization of optical pumping, which we have demonstrated by three examples. State selective absorption imaging can deliver spatially resolved information about the spin state, also for samples at $\mu$K temperatures. Using these methods, we have determined an upper bound for the $^{87}$Sr spin relaxation rate and have found it to be low, as expected. These methods will be necessary tools for the implementation of quantum simulations and quantum computation making use of the $^{87}$Sr nuclear spin.

\begin{acknowledgments}

We gratefully acknowledge support from the Austrian Ministry of Science and Research (BMWF) and the Austrian Science Fund (FWF) through a START grant under project number Y507-N20 as well as support from the European Commission under project number 250072 iSENSE.

\end{acknowledgments}

\bibliographystyle{apsrev}


\begin{thebibliography}{28}
\expandafter\ifx\csname natexlab\endcsname\relax\def\natexlab#1{#1}\fi
\expandafter\ifx\csname bibnamefont\endcsname\relax
  \def\bibnamefont#1{#1}\fi
\expandafter\ifx\csname bibfnamefont\endcsname\relax
  \def\bibfnamefont#1{#1}\fi
\expandafter\ifx\csname citenamefont\endcsname\relax
  \def\citenamefont#1{#1}\fi
\expandafter\ifx\csname url\endcsname\relax
  \def\url#1{\texttt{#1}}\fi
\expandafter\ifx\csname urlprefix\endcsname\relax\def\urlprefix{URL }\fi
\providecommand{\bibinfo}[2]{#2}
\providecommand{\eprint}[2][]{\url{#2}}

\bibitem[{\citenamefont{Cazalilla et~al.}(2009)\citenamefont{Cazalilla, Ho, and
  Ueda}}]{Cazalilla2009ugo}
\bibinfo{author}{\bibfnamefont{M.~A.} \bibnamefont{Cazalilla}},
  \bibinfo{author}{\bibfnamefont{A.}~\bibnamefont{Ho}}, \bibnamefont{and}
  \bibinfo{author}{\bibfnamefont{M.}~\bibnamefont{Ueda}}, \bibinfo{journal}{New
  J. Phys.} \textbf{\bibinfo{volume}{11}}, \bibinfo{pages}{103033}
  (\bibinfo{year}{2009}).

\bibitem[{\citenamefont{Gorshkov et~al.}(2010)\citenamefont{Gorshkov, Hermele,
  Gurarie, Xu, Julienne, Ye, Zoller, Demler, Lukin, and Rey}}]{Gorshkov2010tos}
\bibinfo{author}{\bibfnamefont{A.}~\bibnamefont{Gorshkov}},
  \bibinfo{author}{\bibfnamefont{M.}~\bibnamefont{Hermele}},
  \bibinfo{author}{\bibfnamefont{V.}~\bibnamefont{Gurarie}},
  \bibinfo{author}{\bibfnamefont{C.}~\bibnamefont{Xu}},
  \bibinfo{author}{\bibfnamefont{P.}~\bibnamefont{Julienne}},
  \bibinfo{author}{\bibfnamefont{J.}~\bibnamefont{Ye}},
  \bibinfo{author}{\bibfnamefont{P.}~\bibnamefont{Zoller}},
  \bibinfo{author}{\bibfnamefont{E.}~\bibnamefont{Demler}},
  \bibinfo{author}{\bibfnamefont{M.~D.} \bibnamefont{Lukin}}, \bibnamefont{and}
  \bibinfo{author}{\bibfnamefont{A.~M.} \bibnamefont{Rey}},
  \bibinfo{journal}{Nature Phys.} \textbf{\bibinfo{volume}{6}},
  \bibinfo{pages}{289} (\bibinfo{year}{2010}).

\bibitem[{\citenamefont{Wu et~al.}(2003)\citenamefont{Wu, Hu, and
  Zhang}}]{Wu2003ess}
\bibinfo{author}{\bibfnamefont{C.}~\bibnamefont{Wu}},
  \bibinfo{author}{\bibfnamefont{J.-P.} \bibnamefont{Hu}}, \bibnamefont{and}
  \bibinfo{author}{\bibfnamefont{S.-C.} \bibnamefont{Zhang}},
  \bibinfo{journal}{Phys. Rev. Lett.} \textbf{\bibinfo{volume}{91}},
  \bibinfo{pages}{186402} (\bibinfo{year}{2003}).

\bibitem[{\citenamefont{Wu}(2006)}]{Wu2006hsa}
\bibinfo{author}{\bibfnamefont{C.}~\bibnamefont{Wu}}, \bibinfo{journal}{Mod.
  Phys. Lett. B} \textbf{\bibinfo{volume}{20}}, \bibinfo{pages}{1707}
  (\bibinfo{year}{2006}).

\bibitem[{\citenamefont{Foss-Feig et~al.}(2010)\citenamefont{Foss-Feig,
  Hermele, and Rey}}]{FossFeig2010ptk}
\bibinfo{author}{\bibfnamefont{M.}~\bibnamefont{Foss-Feig}},
  \bibinfo{author}{\bibfnamefont{M.}~\bibnamefont{Hermele}}, \bibnamefont{and}
  \bibinfo{author}{\bibfnamefont{A.~M.} \bibnamefont{Rey}},
  \bibinfo{journal}{Phys. Rev. A} \textbf{\bibinfo{volume}{81}},
  \bibinfo{pages}{051603} (\bibinfo{year}{2010}).

\bibitem[{\citenamefont{Hermele et~al.}(2009)\citenamefont{Hermele, Gurarie,
  and Rey}}]{Hermele2009mio}
\bibinfo{author}{\bibfnamefont{M.}~\bibnamefont{Hermele}},
  \bibinfo{author}{\bibfnamefont{V.}~\bibnamefont{Gurarie}}, \bibnamefont{and}
  \bibinfo{author}{\bibfnamefont{A.~M.} \bibnamefont{Rey}},
  \bibinfo{journal}{Phys. Rev. Lett.} \textbf{\bibinfo{volume}{103}},
  \bibinfo{pages}{135301} (\bibinfo{year}{2009}).

\bibitem[{\citenamefont{Xu}(2010)}]{Xu2010lim}
\bibinfo{author}{\bibfnamefont{C.}~\bibnamefont{Xu}}, \bibinfo{journal}{Phys.
  Rev. B} \textbf{\bibinfo{volume}{81}}, \bibinfo{pages}{144431}
  (\bibinfo{year}{2010}).

\bibitem[{\citenamefont{Hung et~al.}(2011)\citenamefont{Hung, Wang, and
  Wu}}]{Hung2011qmo}
\bibinfo{author}{\bibfnamefont{H.-H.} \bibnamefont{Hung}},
  \bibinfo{author}{\bibfnamefont{Y.}~\bibnamefont{Wang}}, \bibnamefont{and}
  \bibinfo{author}{\bibfnamefont{C.}~\bibnamefont{Wu}},
  \bibinfo{journal}{arXiv:1103.1926}  (\bibinfo{year}{2011}).

\bibitem[{\citenamefont{Gerbier and Dalibard}(2010)}]{Gerbier2009gff}
\bibinfo{author}{\bibfnamefont{F.}~\bibnamefont{Gerbier}} \bibnamefont{and}
  \bibinfo{author}{\bibfnamefont{J.}~\bibnamefont{Dalibard}},
  \bibinfo{journal}{New J. Phys.} \textbf{\bibinfo{volume}{12}},
  \bibinfo{pages}{033007} (\bibinfo{year}{2010}).

\bibitem[{\citenamefont{Cooper}(2011)}]{Cooper2011ofl}
\bibinfo{author}{\bibfnamefont{N.~R.} \bibnamefont{Cooper}},
  \bibinfo{journal}{Phys. Rev. Lett.} \textbf{\bibinfo{volume}{106}},
  \bibinfo{pages}{175301} (\bibinfo{year}{2011}).

\bibitem[{\citenamefont{B{\' e}ri and Cooper}(2011)}]{Beri2011zti}
\bibinfo{author}{\bibfnamefont{B.}~\bibnamefont{B{\' e}ri}} \bibnamefont{and}
  \bibinfo{author}{\bibfnamefont{N.~R.} \bibnamefont{Cooper}},
  \bibinfo{journal}{arXiv:1105.1252}  (\bibinfo{year}{2011}).

\bibitem[{\citenamefont{G\'orecka et~al.}(2011)\citenamefont{G\'orecka,
  Gr\'{e}maud, and Miniatura}}]{Gorecka2011smf}
\bibinfo{author}{\bibfnamefont{A.}~\bibnamefont{G\'orecka}},
  \bibinfo{author}{\bibfnamefont{B.}~\bibnamefont{Gr\'{e}maud}},
  \bibnamefont{and}
  \bibinfo{author}{\bibfnamefont{C.}~\bibnamefont{Miniatura}},
  \bibinfo{journal}{arXiv:1105.3535}  (\bibinfo{year}{2011}).

\bibitem[{\citenamefont{Daley et~al.}(2008)\citenamefont{Daley, Boyd, Ye, and
  Zoller}}]{Daley2008qcw}
\bibinfo{author}{\bibfnamefont{A.~J.} \bibnamefont{Daley}},
  \bibinfo{author}{\bibfnamefont{M.~M.} \bibnamefont{Boyd}},
  \bibinfo{author}{\bibfnamefont{J.}~\bibnamefont{Ye}}, \bibnamefont{and}
  \bibinfo{author}{\bibfnamefont{P.}~\bibnamefont{Zoller}},
  \bibinfo{journal}{Phys. Rev. Lett.} \textbf{\bibinfo{volume}{101}},
  \bibinfo{pages}{170504} (\bibinfo{year}{2008}).

\bibitem[{\citenamefont{Gorshkov et~al.}(2009)\citenamefont{Gorshkov, Rey,
  Daley, Boyd, Ye, Zoller, and Lukin}}]{Gorshkov2009aem}
\bibinfo{author}{\bibfnamefont{A.~V.} \bibnamefont{Gorshkov}},
  \bibinfo{author}{\bibfnamefont{A.~M.} \bibnamefont{Rey}},
  \bibinfo{author}{\bibfnamefont{A.~J.} \bibnamefont{Daley}},
  \bibinfo{author}{\bibfnamefont{M.~M.} \bibnamefont{Boyd}},
  \bibinfo{author}{\bibfnamefont{J.}~\bibnamefont{Ye}},
  \bibinfo{author}{\bibfnamefont{P.}~\bibnamefont{Zoller}}, \bibnamefont{and}
  \bibinfo{author}{\bibfnamefont{M.~D.} \bibnamefont{Lukin}},
  \bibinfo{journal}{Phys. Rev. Lett.} \textbf{\bibinfo{volume}{102}},
  \bibinfo{pages}{110503} (\bibinfo{year}{2009}).

\bibitem[{\citenamefont{Fukuhara et~al.}(2007)\citenamefont{Fukuhara, Takasu,
  Kumakura, and Takahashi}}]{Fukuhara2007dfg}
\bibinfo{author}{\bibfnamefont{T.}~\bibnamefont{Fukuhara}},
  \bibinfo{author}{\bibfnamefont{Y.}~\bibnamefont{Takasu}},
  \bibinfo{author}{\bibfnamefont{M.}~\bibnamefont{Kumakura}}, \bibnamefont{and}
  \bibinfo{author}{\bibfnamefont{Y.}~\bibnamefont{Takahashi}},
  \bibinfo{journal}{Phys. Rev. Lett.} \textbf{\bibinfo{volume}{98}},
  \bibinfo{pages}{030401} (\bibinfo{year}{2007}).

\bibitem[{\citenamefont{Taie et~al.}(2010)\citenamefont{Taie, Takasu, Sugawa,
  Yamazaki, Tsujimoto, Murakami, and Takahashi}}]{Taie2010roa}
\bibinfo{author}{\bibfnamefont{S.}~\bibnamefont{Taie}},
  \bibinfo{author}{\bibfnamefont{Y.}~\bibnamefont{Takasu}},
  \bibinfo{author}{\bibfnamefont{S.}~\bibnamefont{Sugawa}},
  \bibinfo{author}{\bibfnamefont{R.}~\bibnamefont{Yamazaki}},
  \bibinfo{author}{\bibfnamefont{T.}~\bibnamefont{Tsujimoto}},
  \bibinfo{author}{\bibfnamefont{R.}~\bibnamefont{Murakami}}, \bibnamefont{and}
  \bibinfo{author}{\bibfnamefont{Y.}~\bibnamefont{Takahashi}},
  \bibinfo{journal}{Phys. Rev. Lett.} \textbf{\bibinfo{volume}{105}},
  \bibinfo{pages}{190401} (\bibinfo{year}{2010}).

\bibitem[{\citenamefont{DeSalvo et~al.}(2010)\citenamefont{DeSalvo, Yan,
  Mickelson, Martinez~de Escobar, and Killian}}]{DeSalvo2010dfg}
\bibinfo{author}{\bibfnamefont{B.~J.} \bibnamefont{DeSalvo}},
  \bibinfo{author}{\bibfnamefont{M.}~\bibnamefont{Yan}},
  \bibinfo{author}{\bibfnamefont{P.~G.} \bibnamefont{Mickelson}},
  \bibinfo{author}{\bibfnamefont{Y.~N.} \bibnamefont{Martinez~de Escobar}},
  \bibnamefont{and} \bibinfo{author}{\bibfnamefont{T.~C.}
  \bibnamefont{Killian}}, \bibinfo{journal}{Phys. Rev. Lett.}
  \textbf{\bibinfo{volume}{105}}, \bibinfo{pages}{030402}
  (\bibinfo{year}{2010}).

\bibitem[{\citenamefont{Tey et~al.}(2010)\citenamefont{Tey, Stellmer, Grimm,
  and Schreck}}]{Tey2010ddb}
\bibinfo{author}{\bibfnamefont{M.~K.} \bibnamefont{Tey}},
  \bibinfo{author}{\bibfnamefont{S.}~\bibnamefont{Stellmer}},
  \bibinfo{author}{\bibfnamefont{R.}~\bibnamefont{Grimm}}, \bibnamefont{and}
  \bibinfo{author}{\bibfnamefont{F.}~\bibnamefont{Schreck}},
  \bibinfo{journal}{Phys. Rev. A} \textbf{\bibinfo{volume}{82}},
  \bibinfo{pages}{011608(R)} (\bibinfo{year}{2010}).

\bibitem[{\citenamefont{Mukaiyama et~al.}(2003)\citenamefont{Mukaiyama, Katori,
  Ido, Li, and Kuwata-Gonokami}}]{Mukaiyama2003rll}
\bibinfo{author}{\bibfnamefont{T.}~\bibnamefont{Mukaiyama}},
  \bibinfo{author}{\bibfnamefont{H.}~\bibnamefont{Katori}},
  \bibinfo{author}{\bibfnamefont{T.}~\bibnamefont{Ido}},
  \bibinfo{author}{\bibfnamefont{Y.}~\bibnamefont{Li}}, \bibnamefont{and}
  \bibinfo{author}{\bibfnamefont{M.}~\bibnamefont{Kuwata-Gonokami}},
  \bibinfo{journal}{Phys. Rev. Lett.} \textbf{\bibinfo{volume}{90}},
  \bibinfo{pages}{113002} (\bibinfo{year}{2003}).

\bibitem[{\citenamefont{Boyd et~al.}(2007)\citenamefont{Boyd, Zelevinsky,
  Ludlow, Blatt, Zanon-Willette, Foreman, and Ye}}]{Boyd2007nse}
\bibinfo{author}{\bibfnamefont{M.~M.} \bibnamefont{Boyd}},
  \bibinfo{author}{\bibfnamefont{T.}~\bibnamefont{Zelevinsky}},
  \bibinfo{author}{\bibfnamefont{A.~D.} \bibnamefont{Ludlow}},
  \bibinfo{author}{\bibfnamefont{S.}~\bibnamefont{Blatt}},
  \bibinfo{author}{\bibfnamefont{T.}~\bibnamefont{Zanon-Willette}},
  \bibinfo{author}{\bibfnamefont{S.~M.} \bibnamefont{Foreman}},
  \bibnamefont{and} \bibinfo{author}{\bibfnamefont{J.}~\bibnamefont{Ye}},
  \bibinfo{journal}{Phys. Rev. A} \textbf{\bibinfo{volume}{76}},
  \bibinfo{eid}{022510} (\bibinfo{year}{2007}).

\bibitem[{\citenamefont{Gerlach and Stern}(1922)}]{Stern1922den}
\bibinfo{author}{\bibfnamefont{W.}~\bibnamefont{Gerlach}} \bibnamefont{and}
  \bibinfo{author}{\bibfnamefont{O.}~\bibnamefont{Stern}}, \bibinfo{journal}{Z.
  Phys.} \textbf{\bibinfo{volume}{8}}, \bibinfo{pages}{110}
  (\bibinfo{year}{1922}).

\bibitem[{\citenamefont{Stamper-Kurn et~al.}(1998)\citenamefont{Stamper-Kurn,
  Andrews, Chikkatur, Inouye, Miesner, Stenger, and
  Ketterle}}]{StamperKurn1998oco}
\bibinfo{author}{\bibfnamefont{D.~M.} \bibnamefont{Stamper-Kurn}},
  \bibinfo{author}{\bibfnamefont{M.~R.} \bibnamefont{Andrews}},
  \bibinfo{author}{\bibfnamefont{A.~P.} \bibnamefont{Chikkatur}},
  \bibinfo{author}{\bibfnamefont{S.}~\bibnamefont{Inouye}},
  \bibinfo{author}{\bibfnamefont{H.-J.} \bibnamefont{Miesner}},
  \bibinfo{author}{\bibfnamefont{J.}~\bibnamefont{Stenger}}, \bibnamefont{and}
  \bibinfo{author}{\bibfnamefont{W.}~\bibnamefont{Ketterle}},
  \bibinfo{journal}{Phys. Rev. Lett.} \textbf{\bibinfo{volume}{80}},
  \bibinfo{pages}{2027} (\bibinfo{year}{1998}).

\bibitem[{\citenamefont{Sleator et~al.}(1992)\citenamefont{Sleator, Pfau,
  Balykin, Carnal, and Mlynek}}]{Sleator1992edo}
\bibinfo{author}{\bibfnamefont{T.}~\bibnamefont{Sleator}},
  \bibinfo{author}{\bibfnamefont{T.}~\bibnamefont{Pfau}},
  \bibinfo{author}{\bibfnamefont{V.}~\bibnamefont{Balykin}},
  \bibinfo{author}{\bibfnamefont{O.}~\bibnamefont{Carnal}}, \bibnamefont{and}
  \bibinfo{author}{\bibfnamefont{J.}~\bibnamefont{Mlynek}},
  \bibinfo{journal}{Phys. Rev. Lett.} \textbf{\bibinfo{volume}{68}},
  \bibinfo{pages}{1996} (\bibinfo{year}{1992}).

\bibitem[{\citenamefont{Metcalf and van~der Straten}(1999)}]{Metcalf1999book}
\bibinfo{author}{\bibfnamefont{H.~J.} \bibnamefont{Metcalf}} \bibnamefont{and}
  \bibinfo{author}{\bibfnamefont{P.}~\bibnamefont{van~der Straten}},
  \emph{\bibinfo{title}{Laser Cooling and Trapping}}
  (\bibinfo{publisher}{Springer, New York}, \bibinfo{year}{1999}).

\bibitem[{End({\natexlab{a}})}]{EndnoteOSGBeamIntensityVariation}
\bibinfo{note}{Assuming OSG beam intensities different by $\pm$20\% from the
  measured values did not change the outcome of the simulation significantly
  after fitting OSG beam waists and positions again.}

\bibitem[{\citenamefont{Matthews et~al.}(1998)\citenamefont{Matthews, Hall,
  Jin, Ensher, Wieman, Cornell, Dalfovo, Minniti, and
  Stringari}}]{Matthews1998dro}
\bibinfo{author}{\bibfnamefont{M.~R.} \bibnamefont{Matthews}},
  \bibinfo{author}{\bibfnamefont{D.~S.} \bibnamefont{Hall}},
  \bibinfo{author}{\bibfnamefont{D.~S.} \bibnamefont{Jin}},
  \bibinfo{author}{\bibfnamefont{J.~R.} \bibnamefont{Ensher}},
  \bibinfo{author}{\bibfnamefont{C.~E.} \bibnamefont{Wieman}},
  \bibinfo{author}{\bibfnamefont{E.~A.} \bibnamefont{Cornell}},
  \bibinfo{author}{\bibfnamefont{F.}~\bibnamefont{Dalfovo}},
  \bibinfo{author}{\bibfnamefont{C.}~\bibnamefont{Minniti}}, \bibnamefont{and}
  \bibinfo{author}{\bibfnamefont{S.}~\bibnamefont{Stringari}},
  \bibinfo{journal}{Phys. Rev. Lett.} \textbf{\bibinfo{volume}{81}},
  \bibinfo{pages}{243} (\bibinfo{year}{1998}).

\bibitem[{End({\natexlab{b}})}]{EndnoteAbsImagingSimulationDetails}
\bibinfo{note}{The simulation determines the average number of photons
  scattered by an atom in a certain $m_F$ state for a distribution of atoms
  corresponding to the one used in our experiment. It takes into account the
  Zeeman splitting of the excited state, the Doppler shift, acceleration of
  atoms by photon absorption and emission, and optical pumping. The result of
  the simulation is that the number of photons absorbed by an atom initially in
  a certain $m_F$ state is to within 10\% proportional to the line strength of
  the transition corresponding to this $m_F$ state.}

\bibitem[{Jul()}]{JuliennePrivateComm}
\bibinfo{note}{P.~S.~Julienne, private communication.}

\end{thebibliography}

\end{document}